# Performance of RPL in Healthcare Wireless Sensor Network

**Bassam Al-Shargabi[1], Mohammed Aleswid[2]**
[1]Department of Computer Science, Middle East University, Amman, Jordan
[1]bshargabi@meu.edu.jo, bassam20_152@yahoo.com
[2]mohmad.aliswid@hotmail.com

## ABSTRACT

The new advances of the Internet of Things (IoT) technology can be utilized to promote service delivery in several real-life applications such as the healthcare systems. The Routing Protocol for Low Power and Loss Network (RPL) is a routing protocol designed to serve as a proper routing protocol for packets in Wireless Sensor Networks (WSN). Among the most prominent issues exist in the RPL protocol are packet loss within the WSN and sensors power consumption especially in healthcare WSNs. Multiple Objective Functions (OF) in RPL intended to find the routes from source nodes to a destination node. This paper presents an evaluation to discover which OF is more efficient for a WSN in a healthcare scenario where the Packet Delivery Ratio (PDR) of WSN and the sensors' power consumption are prominent concerns. Expected transmission Count (ETX) and Objective Function Zero (OF0) of RPL were examined in various network densities and network topologies such as the grid and random topology. The simulation outcomes revealed that the OF0 is more efficient regarding the PDR and power consumption compared to the ETX in random topology.

## 1. INTRODUCTION

The expanded use of smart technologies such Big Data, Cloud computing, and Internet of Things (IoT) that makes users more dependent on computers and networks. Newly, the IoT has impacted every aspect of human life and industries such as healthcare, smart grid, and smart homes which all achieved through the Wireless Sensor Network (WSN) [1]. The WSN can be outlined as a set of sensors that are employed in the sense or monitor particular physical or biochemical aspects without the involvement of human[2],[3].

The routing in WSN relies on RPL protocol, where RPL protocol is designed to be an inter-operable and simple protocol for the interconnected IoT sensors or devices (resource-constrained) to be exploited in manufacturing, hospitals, and smart homes [4]. RPL forms a topology comparable to a tree where each sensor or node in the network has an been assigned with a rank, in which it grows as the nodes move faraway from the root node. The RPL specifies the route based on routing metrics and restrictions that should be applied to attain specific purposes which can be achieved by the RPL through the use of OFs. For instance, the OF may designate with the aim of finding the shortest path where the constraint is associated with the node power consumption [5],[ 6].

In RPL OF0 is intended to attain the nearest grounded root where that could be accomplished only if the node rank is determined by the degree its adjacency to the root node. This demand can be estimated with the other needs of having other path options, which can be realized by improving the node rank [7]. Another objective function is the EXT, where this OF relies on the number of the retransmissions ratio of the packet to be delivered successfully within WSN. The RPL supports the application of OF to create route paths that can be controlled by a routing metric. This designation defined by ETXOF to reduces the ETX. The computation of the path is based on ETXOF where it occurs in minimum-ETX paths to the DAG roots from the nodes, where such path can lead to reducing the packet transmissions times from nodes in the WSN to the DAG root [8],[9]. The ETX is viewed as a link measure for predicting the transmissions of the packet to be delivered to destination through acquiring the most suitable path and anticipating the retransmissions number for the packet to be received.

To decide which OF is more efficient when implementing the RPL protocol for the PDR metric, which can be estimated the number of successfully transmitted packets by the root node where it is also correlated to the number transmitted packets by clients. The higher the percentage of PDR means the efficient routing protocol regarding the delivered packet ratio[10], [11]. Moreover, the power consumption metric must be considered, where IoT network or WSN devices are resources constrained devices such as sensors in terms of power or processing. Therefore, the long lifetime nodes in WSN have directed to finding out distinct extents for using a specific implementation of RPL to consider preserving the nodes' consumption of power[12],[13]. To improve the sensor node's power endurance, protocols must be efficient in term of energy through performing prior actions by assessing and foretelling the nodes power consumption degree [14],[15].

In a healthcare system where the IoT or WSN is a major component, choosing which objective function to be used is a major dilemma. Therefore, this paper presents an experimental evaluation of ETX and OF0 objective function of RPL to evaluate their effectiveness regarding power consumption and PDR in a healthcare scenario under different topologies.

The rest of the paper organized as follows: section 2 outlines the most recent related work. The elaboration of the performance evaluation is presented in section 3. Results and discussion are outlined in section 4. Finally, the conclusion and future work are drawn in Section 5

## 2. Related Work

Many approaches have been proposed to tackle the issue of ensuring the data delivery within IoT network with consideration of the limited resources of IoT devices. This section introduces the most recent related works to the use of RPL protocol as presented below:

An evaluation of RPL was conducted in [16], where the evaluation was based on network Latency, loss of beam, and sensors power. The authors used 80 nodes in their experiments, where OF0 was examined through including the counting hops and ETX utilized to determine the optimum routes. As a result, the ETX is outperformed the 0F0 because of the confluence of network time, enhanced traffic, and nodes consumption for power except that the high number of the retransmitted packet is considered as an obstacle[17].

A comparative study was presented in [18], ,which is primarily based on MRHOF and OF as OFs, where they conducted simulation relies on 30 nodes and was implemented using random among other topologies to measure the implemented OFs to show nodes energy consumption rate and also the PDR. Their outcomes revealed that MRHOF shows comparable results with OF0 with regards to the PDR and power consumption, even as in [19] the OF0 and MRHOF were also implemented but for the OF0, the nodes can be selected based on the bare minimum number of hops to the destination. While in MRHOF the parent node selected based on the reliability of the delivered packet. Different performance analysis of RPL was conducted in [20], on the same two OFs, where they analyzed used constant topology and random adjustable networks of 80 nodes with 3 transmitting bandss and the result suggests that OF0 is more efficient regarding the nodes power consumption. In [21], an approach was implemented RPL on fixed and mobile nodes to predict power consumption durability of sensor nodes by using a multiple metrics which include the radio obligation cycle, number of hops, and power mode for each node in the WSN.

An assessment of the performance of multi-instances of RPL via the use of two OFs in [22]. The assessment carried out the implementation of RPL using single and multi-instance regarding PDR, routing tree convergence, and latency as factors for the overall performance. Their simulations were based on two data traffic types labeled as ordinary and crucial data traffic and also based on three varied RX (70%, 85%, 100%.) and concerning the routing tree convergence metric, the outcome revealed that routing tree convergence time was impacted by the use of a multi-instance of RPL compared to a single instance of RPL, this due to the fact that each sensor node has to enroll in the each DAGs which is reflected on the convergence time to complete the DAGs construction. Furthermore, the usage of multi-instance RPL has led to higher latency and PDR as compared to single instance RPL. Besides, in [23], multi-instance of RPL with a cooperative approach among times named (C-RPL) where the multi-instance of RPL used to control the power consumption of nodes with WSN with consideration network features also the used OF for each instance of RPL. A major feature of using the C-RPL is the "collation", where it is composed of more than one instance with a shared association between nodes to enhance their utilities. Also, in C-RPL an equity evaluation for networks to manipulate the trade-off among other performance indicators of the network compared to the power consumption factor. The C-RPL was evaluated and examined against standard RPL with different data traffics. The evaluation is based on implementing the four RPL types RPL and C-RPL. The outcome shown that C-RPL will generate instances successfully based on the implemented OF and comply to the conditions of the network. Moreover, the C-RPL proved to be more efficient regarding the power consumption due to the nature of C-RPL in adjusting the number of instances created according to the network densities[24],[23].

## 3. PERFORMANCE EVALUATION

To assess the performance of RPL based on the OF0 and ETX OFs on a WSN healthcare scenario regarding two important factors such as the power consumption and the PDR, along with studying the effects of network topology to be implemented. The implementation was carried out in Cooja simulation where 20, 40, 60, 80, 100 sensors will be placed in different network densities such as 100% and 80% on different topologies and also based on the sending time interval that helps us determine the operations that will take place inside the emulator as the sensors types it will be divided into high critical, critical, low critical (Periodic), and room sensors such as temperature inside a hospital.

### 3.1 Simulation and Network Setup

In this paper, we set up the network using one sink node with two different topologies along with nodes distribution in 1000 meters squared area with placing the at the center of the network. The implementation of RPL based on OF0 and ETX through setting the experiments with different network densities. The network also designed using a varied number of nodes where the RPL network might contain (20, 40, 60, 80 and 100 nodes) with different data traffic specifications for the nodes along with the sink node. in addition, we used a varied

RX value (80% and 100%). PDR and power consumption are the main factors to evaluate RPL implementation based on ETX and OF0. We used the main default RPL parameters as in [18],[22] as shown in Table 1. Along with different values for sending interval time for data traffics' and packet size for the designated healthcare scenarios are shown in table 2. For example, blood oxygen, body temperature, blood pressure, and heart rate sensor data are taken every 5 mints, while the temperature sensor of the room is taken every 1 hour for inpatient rooms while other sensors installed in Intensive Care Unit(ICU) have higher priority; that's why the sending interval of such sensors are between 10 and 20 seconds.

**Table 1.** Parameters used in the Simulation

| Parameters | Value |
|---|---|
| OF | OF0, ETX |
| TX Ratio | 80-100% |
| TX Range | 100m |
| Topologies | Random, Grid |
| Simulation Time | 900 second |
| squared area | 1000 meters |

**Table 2.** The data traffic types and sending intervals

| Traffic Type | Sending Interval |
|---|---|
| High-critical | Average of 10 seconds |
| Critical | Average of 20 seconds |
| Low-critical (periodic) | Every 5 minutes |
| Temperature | Average 60 mint |

### 3.2 Performance Metrics

To evaluate the RPL protocol implementation in a healthcare scenario based on OF0 and ETX as OFs with regarding the power consumption and PDR as performance measures.

### 3.3 Network Topologies

To evaluate OF0 and ETX OFs of RPL in a healthcare scenario. Where the RPL advocates some types of application requirements in the course of using many OFs, with regarding nodes number where it varies from 20 to 100 nodes where the nodes are distributed around the sink node, along with the varied density of the network. Another factor to evaluate RPL protocol is the network topology, where two topologies were considered. The first topology is the random topology, wherein the distribution of nodes were located in a different network densities of (20 -100 nodes) were distributed on the base of sending time interval where each 20 nodes will be assigned with unique sending time interval as in healthcare scenario as described in table2, where 100 contracts were distributed and there is one contract to collect information. As shown in Figure 1, the high critical data traffic nodes are yellow colored, the critical data traffic nodes are represented by purple color, the low critical(Periodic) data traffic are turquoise colored, the temperature data traffic are blue colored, and finally, sink node was represented using green color.

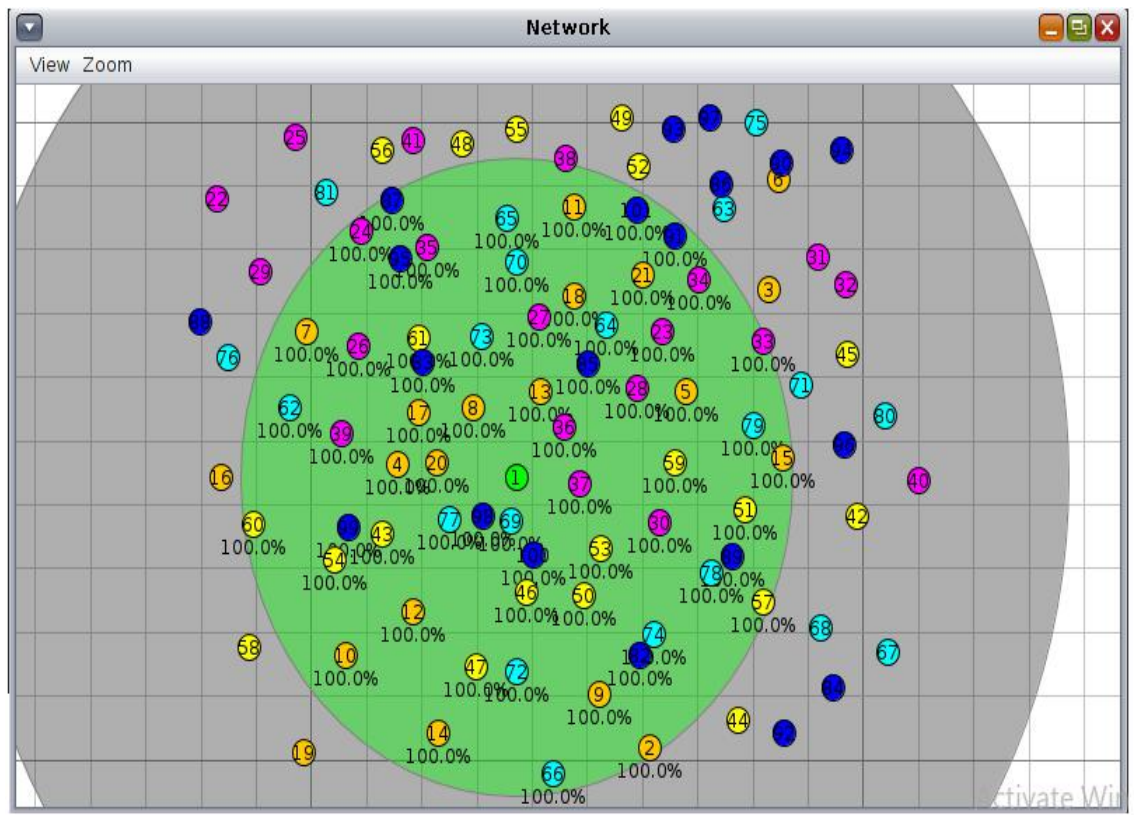

**Figure1:** Random Topology

The grid topology is the second topology, where the nodes distribution allows the communication between nodes to reach the sink, while the network edge nodes handles the transfer of data between nodes in a faster manner, which leads to a drop in energy consumption. In our experiments as shown in Figure 2, the high critical data traffic nodes are yellow colored, the critical data traffic nodes are represented by Turquoise colored, the temperature data traffic is blue colored, and finally, the green-colored node represents the sink node.

## 4. RESULT AND DISCUSSION

This section presents the result and discussion of conducted experiments for evaluating the RPL based on the data gathered via the Cooja simulator. Assessing the OF0 and ETX is main the goal of these experiments . The assessment is based on PDR and power consumption as performance measures or factors. Experiments were conducted on varied numbers of nodes and on a varied topologies to asses its influence of such factors of RPL performance.

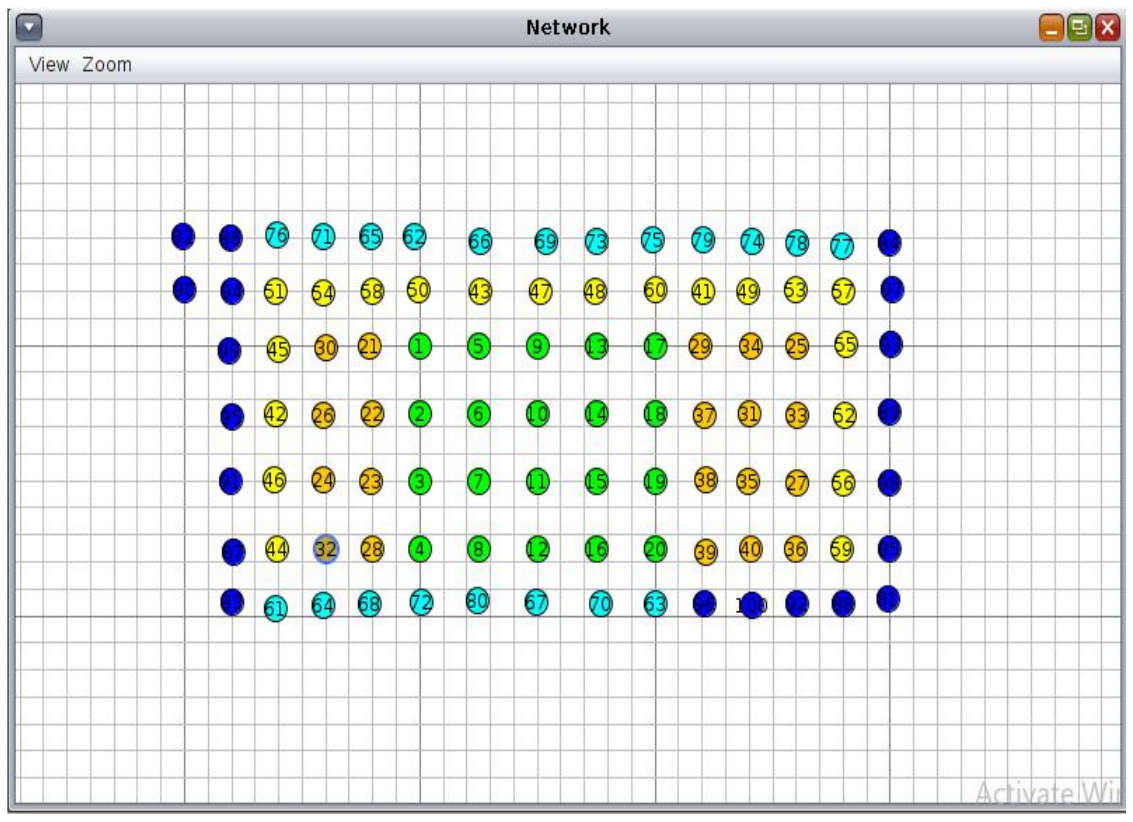

**Figure 2:** Grid Topology

### .4.1 RPL Performance Based On OF0

The experiments were set up to be used with varied network densities (20, 30, 40, and 100), also using the grid and random topologies to assess the performance of RPL based on OF0 with different values of RX (80, and100%) to check the RPL performance regarding the power consumption and PDR.

As shown in Figure 3 and Figure 4, the behavior RPL based on OF0 regarding the ratio of PDR with varied RX for the random and grid topology, as noticed the PDR ratio improved if RX values grow. Moreover, as shown in figure 3, the PDR ratio reaches 98% for the RX equal 100% in random topology compared to the 97% PDR ratio for the grid topology. Figure 4, shows that the PDR reached almost 95 % for RX 80% in random topology compared to the 92% PDR ratio in a grid topology. This means we can select the random topology with RX 100% as an alternative of RX80% for the reason that RPL offers an improved PDR with a percentage of 98%. The rationale behind these results comes from that the RX value is not altered after 80 where it become adequate to deliver the many packets of the LLN.

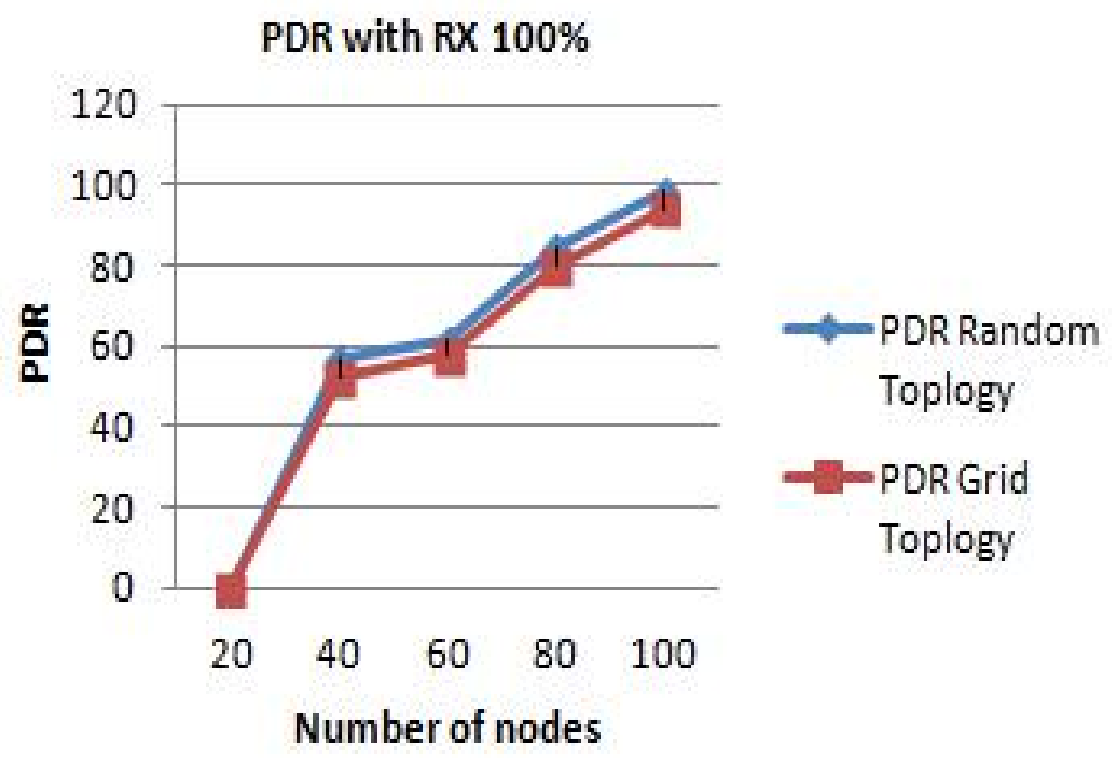

**Figure 3:** PDR ratio with RX 100%

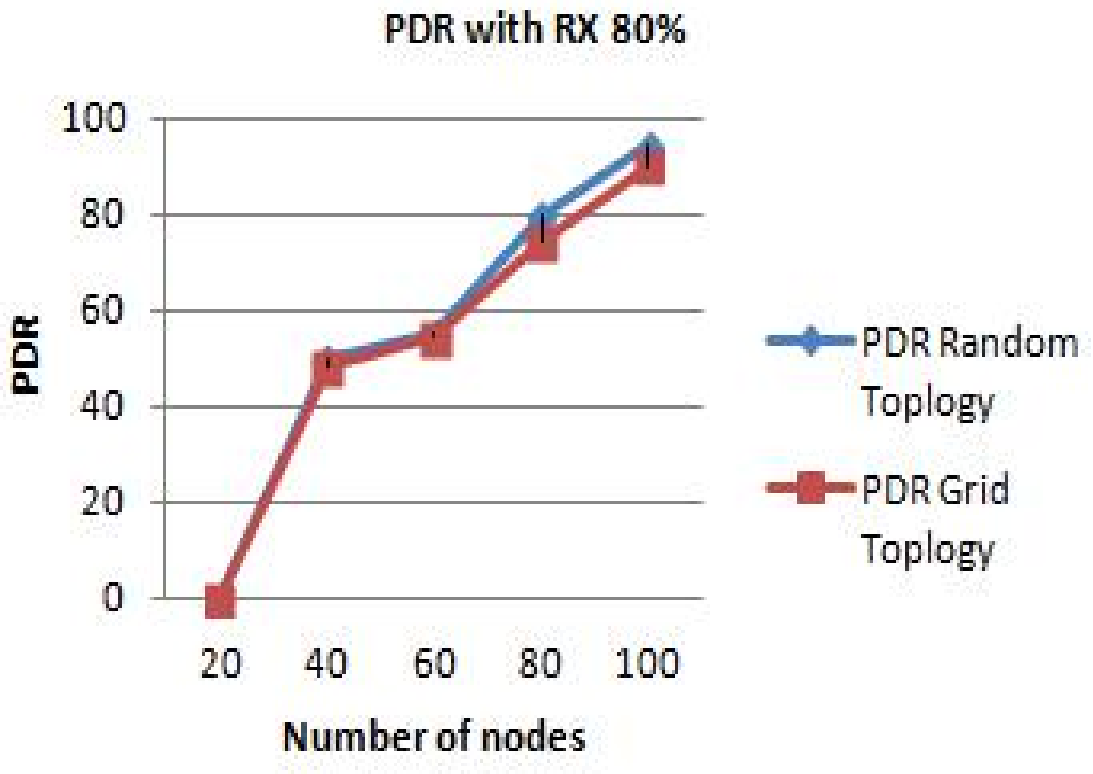

**Figure 4:** PDR ratio with RX 80%

As shown in Figure 5 and Figure 6, the assessment of RPL implementation based on OF0 regarding the consumption of power for nodes based on RX values in random topology, it was noted that the average of nodes power consumption with the RX 100% reached 1.4% as compared to the RX 80% where it reached 1.6.%. The rationale behind these results comes from that the RX value has not been perceived yet at RX 82%, which is enough to preserve energy consumption. Similar results were obtained for the grid topology as well with approximately 1.4% of the power consumption.

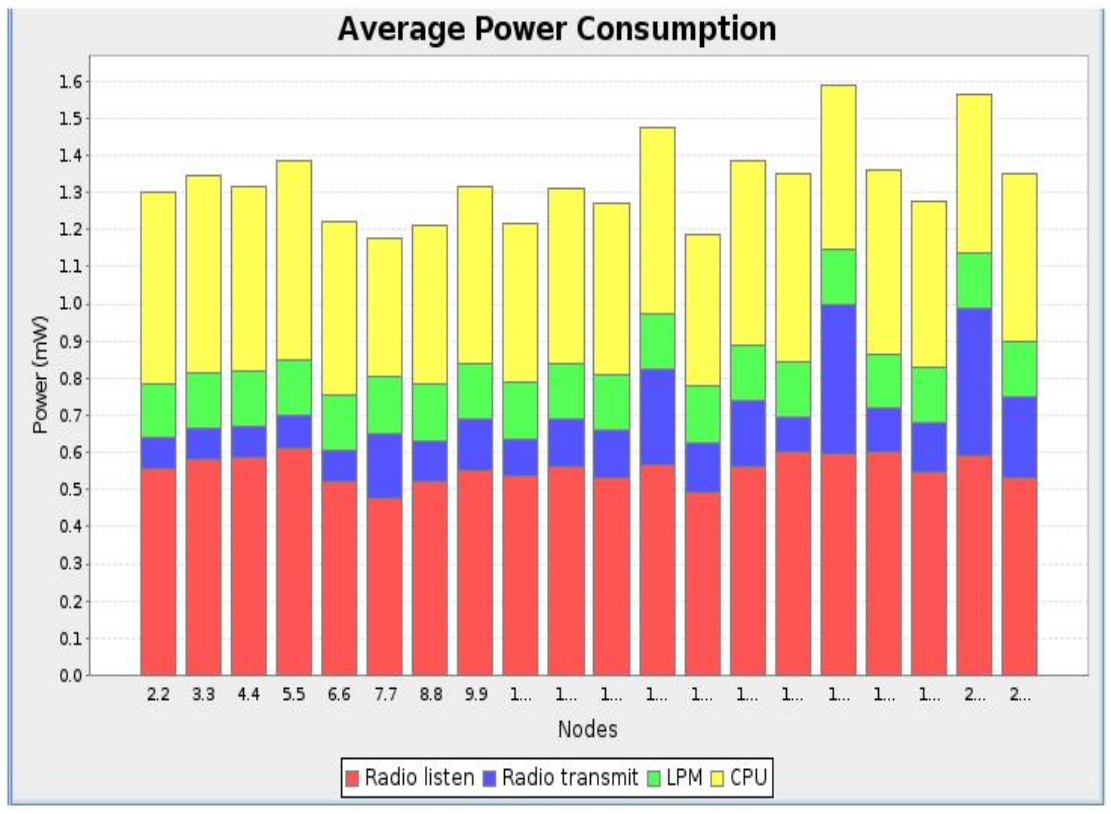

**Figure 5:** Power Consumption with RX 100%

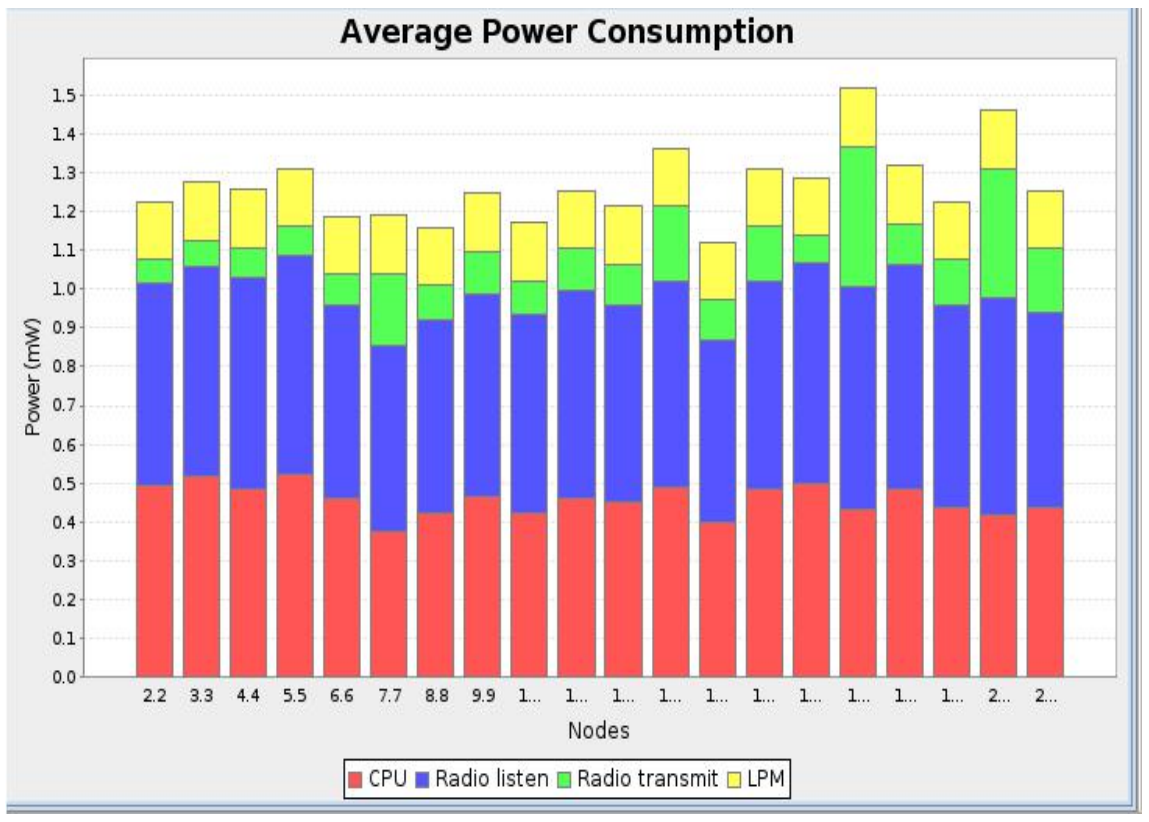

**Figure 6:** Power Consumption with RX 80%

### 4.2 RPL Performance Based On ETX

The experiments were set up to be used with varied different network densities (20, 30, 40, and 100), and also using the grid and random topologies to assess the performance of RPL based on ETX with for different values for RX values (80, and100%) to show its effects on the performance of RPL regarding PDR ratio and power consumption based on ETX objective function. As illustrated in figure 7. The PDR ratio for RX 100% in random topology reached 95 % compared to 92% of PDR in grid topology for 100 nodes. Also, as shown in figure 8 the PDR ration for RX 80% in random topology almost reached 90 % compared to 88 % of PDR in grid topology for 100 nodes in the network.

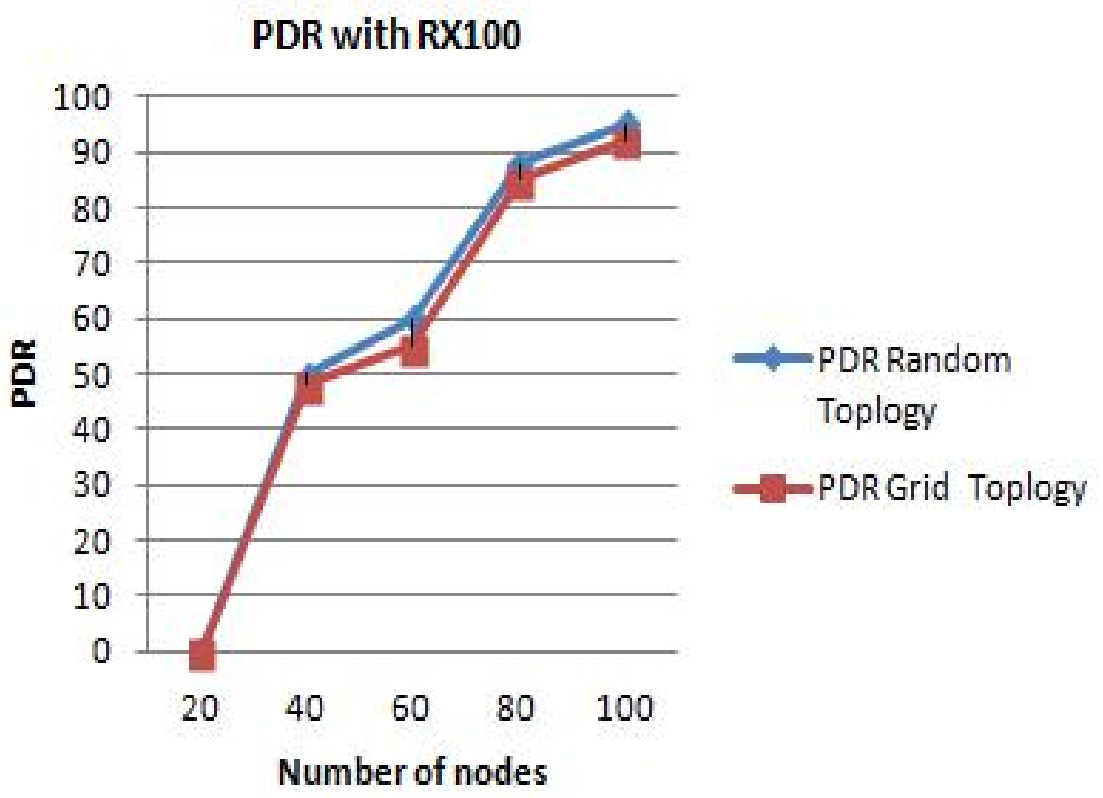

**Figure** 7: PDR Ratio with RX 100%

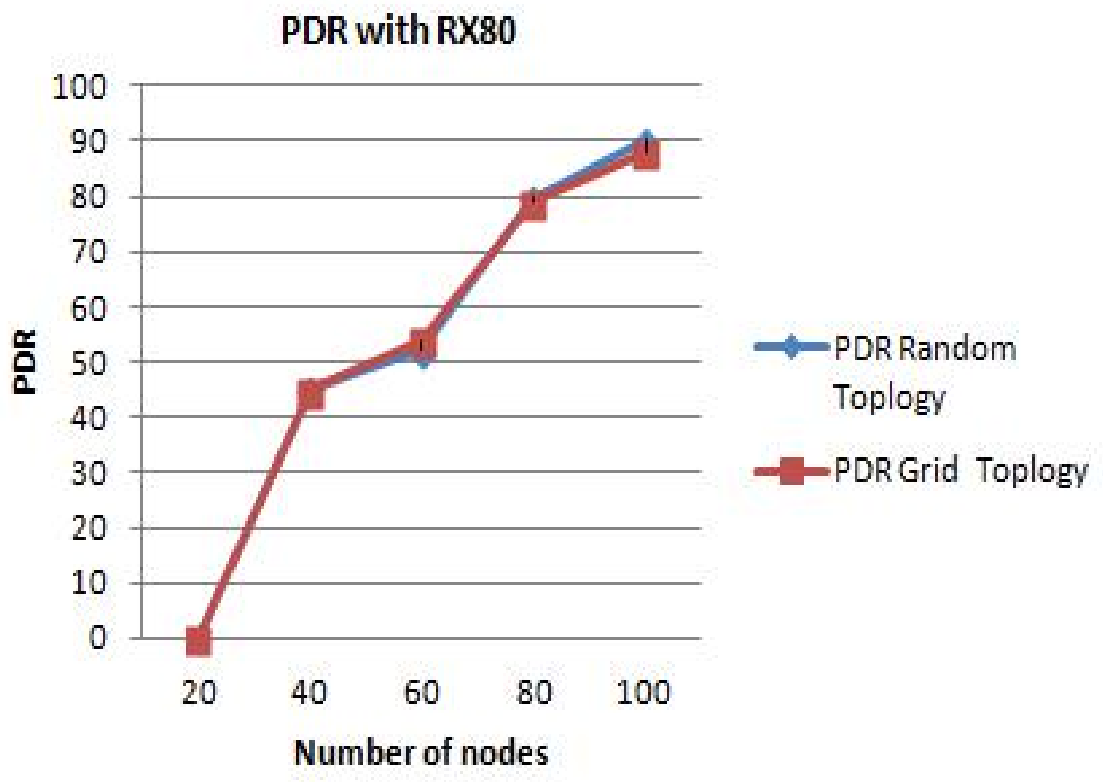

**Figure 8:** PDR Ratio with RX 80%

However, the assessment of RPL for the ETX objective function regarding an important factor in healthcare scenario which is the power consumption, as shown in figure 9, where it reveals the power consumption with varied RX values in random topology, it was noticed that the power consumption percentage has dropped while the RX values have been raised as the average consumption of power. The result showed that with RX 100% we accomplished a result of 1.3% compared to 1.4% with RX 80%. On the other hand, as shown in figure 10, the same result appears for power consumption in random and grid topology which is about 1.4% at RX equals 80%. The rationale behind these results comes from that the RX value has not been perceived yet at RX 82%, which is enough to conserve the power of sensors.

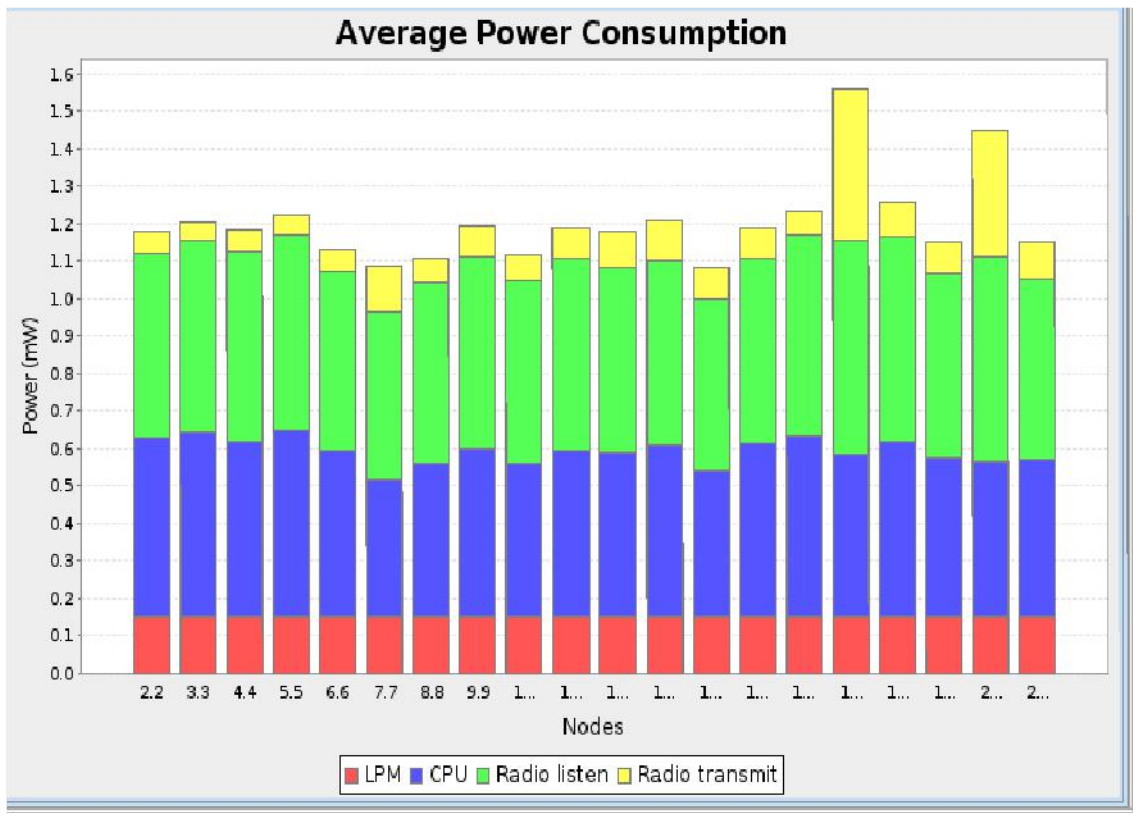

**Figure 9:** Power Consumption with RX 100%

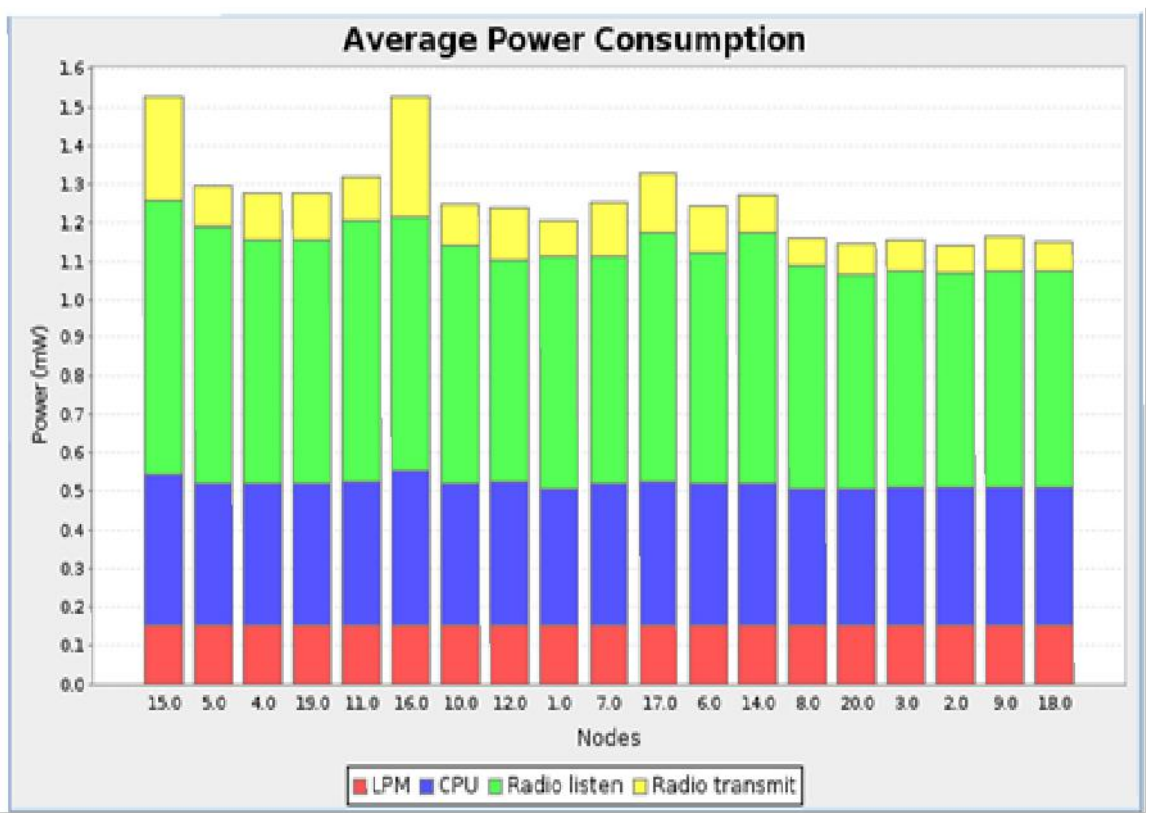

**Figure 10:** Power Consumption with RX 80%

### 4.3 Discussion

To decided which objective function to be used in the proposed healthcare scenario for implementing RPL in WSN with two determining factors such as PDR and power consumption. The two objective OF0 and ETX of RPL were implemented to prove which one is more effective to be used in healthcare WSN. First, the PDR factor, the experimental results in a random topology with RX 100% shows that the average PDR for OF0 is around 98% compared to the average PDR for ETX is reached 95%. Furthermore, if grid topology used instead of random the results showed that the average of PDR for OF0 is almost 97% and the average of PDR for ETX almost reached 92%. On the other hand, the PDR has shown a good PDR ration for OF0 compared to ETX due to the differences in network densities for both topologies.

The second factor for assessing the RPL implementation using OF0 and ETX OFs regarding the nodes power consumption, the results demonstrated that the consumption power for the OF0 reached 1.4% compared to 1.3% of ETX in random topology with RX 100%. In addition, the power consumption average of OF0 was 1.5% in a grid topology compared to 1.4% of ETX. On the other hand, the results showed that the power consumption rate of OF0 reached 1.6% as compared to ETX where it reached 1.4% in random topology with RX 80%. Comparable power consumption rate was noticed of both OF0 and ETX on a grid topology. Certainly, as noted from the result a steady consumption of the power for OF0 and ETX with RX 80%. The simulation outcomes also shown the OF0 drains more power compared to the ETX, but the optimum power consumption for OFs at network density of 100 nodes. Additionally, we have established that original RPL gives comparable results for the PDR for the two OFs in light network densities where the OF0 is more efficient compared to the ETX**.**

## 5. CONCLUSION AND FUTURE WORK

In this paper, we conducted a performance evaluation of implementing RPL relying on OF0 and ETX objective functions in the WSN healthcare scenario to determine which objective function is more effective to meet the specifications of WSN in healthcare with regarding two primary factors as the power consumption and PDR. Simulation experiments were conducted on random and grid topologies with varied RX. These experiments were implemented throughout a particular number of nodes along with various network densities. The outcomes of experiments revealed that the OF0 is more efficient regarding the PDR with the comparable rate on power consumption as compared to ETX. Accordingly, the design of a WSN in healthcare especially in ICU based o the implementation of the RPL protocol would be implemented based on OF0 rather than ETX where the PDR rate must be high because if it was low, patients in ICU might face a high risk or death. Moreover, in this paper, the implementation was based on one instance of RPL, as future work, we intend to investigate the use of multi-instance RPL.

## ACKNOWLEDGMENT

The authors are grateful to the Middle East University, Amman, Jordan for the financial support granted to cover the publication fee of this research article.